\documentclass[aps,prl,twocolumn,notitlepage]{revtex4-1}

\usepackage{amsmath}
\usepackage{amssymb}
\usepackage{graphicx}
\usepackage{epstopdf}

\begin{document}
\title{Interaction Between a Domain Wall and Spin Supercurrent in Easy-cone Magnets}

\author{Se Kwon Kim}
\author{Yaroslav Tserkovnyak}
\affiliation{Department of Physics and Astronomy, University of California, Los Angeles, California 90095, USA}

\begin{abstract}
A domain wall and spin supercurrent can coexist in magnets with easy-cone anisotropy owing to simultaneous spontaneous breaking of Z$_2$ and U(1) symmetries. Their interaction is theoretically investigated in quasi one-dimensional ferromagnets within the Landau-Lifshitz-Gilbert phenomenology. Specifically, we show that spin supercurrent can exert the torque on a domain wall and thereby drive it. We also show, as a reciprocal phenomenon, a field-induced motion of a domain wall can generate spin supercurrent. 
\end{abstract}

\date{\today}
\maketitle

\emph{Introduction.}|Spins in magnets see the crystal lattice through overlap of electron orbitals, which engenders anisotropy energy. In particular, crystal lattices with a single axis of high symmetry, e.g., hexagonal crystals with the axis of sixfold rotational symmetry, endow magnets with uniaxial anisotropy \cite{*[][{, and references therein.}] Kittel}. Uniaxial anisotropy energy is invariant under two operations on spins: the time reversal and the rotations of spins around the axis, which can be characterized by discrete Z$_2$ and continuous U(1) symmetries, respectively.

When the symmetry axis is easy axis, there are two ground states, in which all the spins are either parallel or antiparallel to the axis. The ground states break the Z$_2$ symmetry, but respect the U(1) symmetry. Spontaneous breaking of the discrete symmetry in a continuous field theory entails a domain wall, which is a topological soliton that smoothly interpolates two distinct ground states \cite{*[][{, and references therein.}] Kardar}. Such domain walls in easy-axis magnets have been extensively investigated \cite{*[][{, and references therein.}] KosevichPR1990} due to a fundamental interest as well as practical motivations exemplified by the racetrack memory \cite{ParkinScience2008}. One of the main results of these studies is a collection of various means to drive a domain wall, which includes a magnetic field \cite{SchryerJAP1974} and a spin-polarized electric current \cite{SlonczewskiJMMM1996, *BergerPRB1996}.

When the symmetry axis is hard direction for spins, there are continuously degenerate ground states: uniform spin states in the easy plane perpendicular to the symmetry axis. The ground states break the U(1) symmetry while maintaining the Z$_2$ symmetry. In a classical field theory, a continuous symmetry of the system implies the existence of a conserved quantity according to Noether's theorem \cite{Kardar}. For easy-plane magnets, the conserved quantity is the spin angular momentum projected onto the symmetry axis. In particular, when the broken symmetry is U(1), the conserved quantity can be transported in the form of superfluid. Easy-plane magnets thus can support superfluid spin transport, which is realized by spiraling spin texture within the easy plane \cite{*[][{, and references therein.}] SoninAP2010}. Spin superfluidity has been gaining attention in spintronics as an efficient spin-transport channel owing to its slower decaying than spin transport by quasiparticles such as magnons \cite{KonigPRL2001, *ChenPRB2014, *ChenPRL2015, TakeiPRL2014, *TakeiPRL2015}. 

\begin{figure}
\includegraphics[width=0.9 \columnwidth]{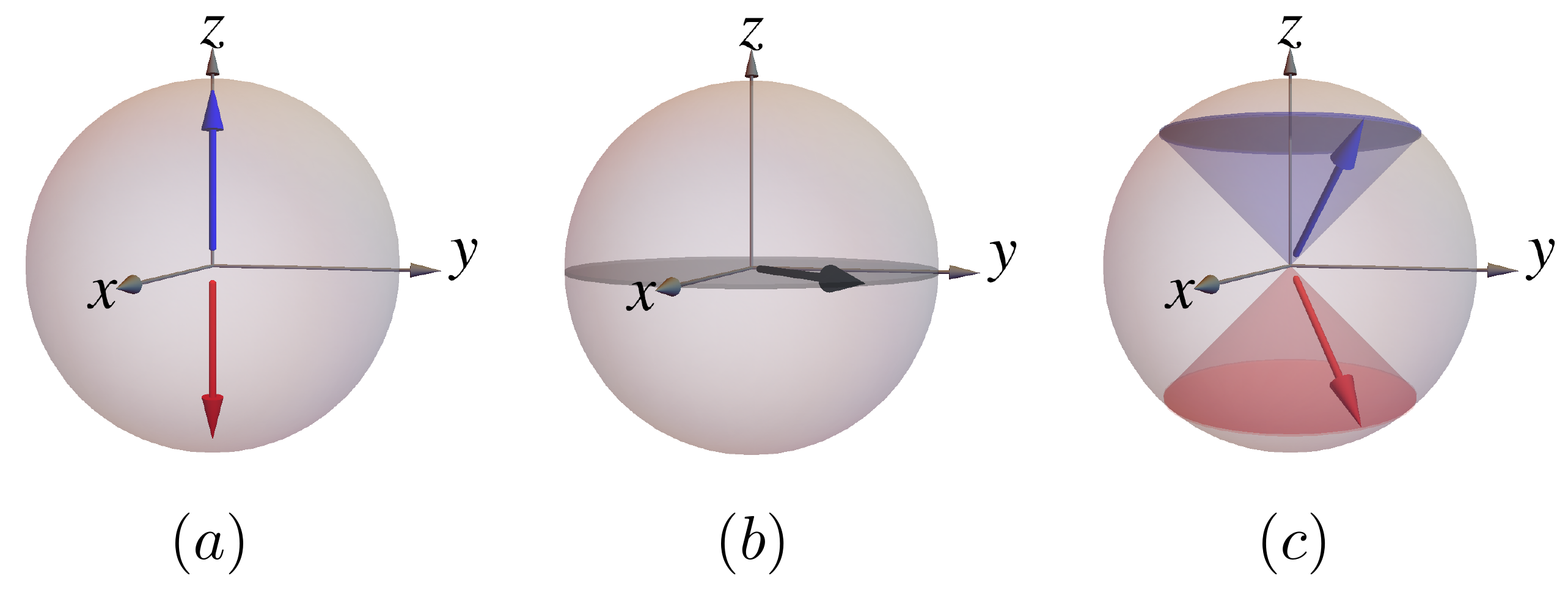}
\caption{(color online) (a) Two arrows represent two ground states of easy-axis magnets. (b) The unit circle in the $xy$ plane represents continuously degenerate ground states of easy-plane magnets; one exemplary spin direction is shown as an arrow. (c) Two cones represent the ground-state manifold of easy-cone magnets; two example spin directions are depicted as arrows.}
\label{fig:fig1}
\end{figure}

\begin{figure}
\includegraphics[width=\columnwidth]{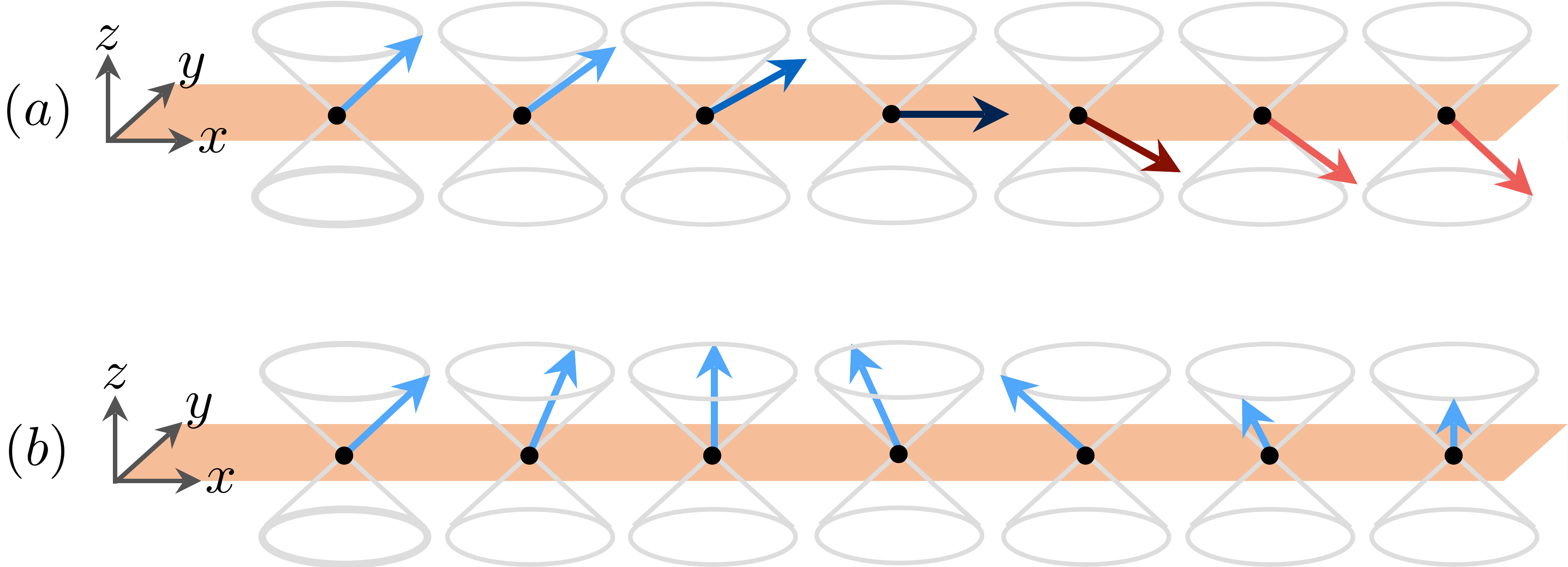}
\caption{(color online) Schematic illustrations of (a) a domain wall and (b) spiraling spin texture within the upper cone, which carries finite spin supercurrent.}
\label{fig:fig2}
\end{figure}

Some magnetic systems have uniaxial anisotropy that is neither easy-axis nor easy-plane anisotropy. Examples of such systems include bilayer Co/Pt \cite{DienyEPL1994, *StampsJAP1997, *FromterPRL2008, *SticklerPRB2011}, multilayer Ta/Co$_{60}$Fe$_{20}$B$_{20}$/MgO \cite{ShawIEEE2015}, and hexagonal compound HoMn$_6$Sn$_4$Ge$_2$ \cite{VenturiniJMMM2009} under their favorable conditions, some of which have been theoretically and experimentally investigated for a memory unit owing to the ease of switching \cite{MatsumotoAPE2015, ShawIEEE2015}. The ground states of those magnets are uniform spin states that tilt away from the symmetry axis. The ground states thereby form two disconnected cones on the unit sphere, which are referred to as easy cones. See Fig.~\ref{fig:fig1} for schematic illustrations of the ground states in uniaxial magnets for comparison of easy-axis, easy-plane, and easy-cone anisotropy. The ground states in easy-cone magnets break both the Z$_2$ and U(1) symmetries, whereby a domain wall and spin superfluidity can coexist. See Fig.~\ref{fig:fig2} for schematic illustrations of a domain wall  and spiraling spin texture carrying finite spin supercurrent. In this Letter, we theoretically study the interaction of a domain wall and spin superfluidity in these systems within the Landau-Lifshitz-Gilbert treatment. Specifically, we show that a domain wall can be driven by spin supercurrent by identifying a spin-transfer torque from the spin supercurrent to the domain wall. We also study its reciprocal phenomenon that spin supercurrent can be generated by the field-induced motion of a domain wall. We conclude the Letter by discussing other possible consequences of the coexistence of a domain wall and spin superfluidity.

\begin{figure}
\includegraphics[width=\columnwidth]{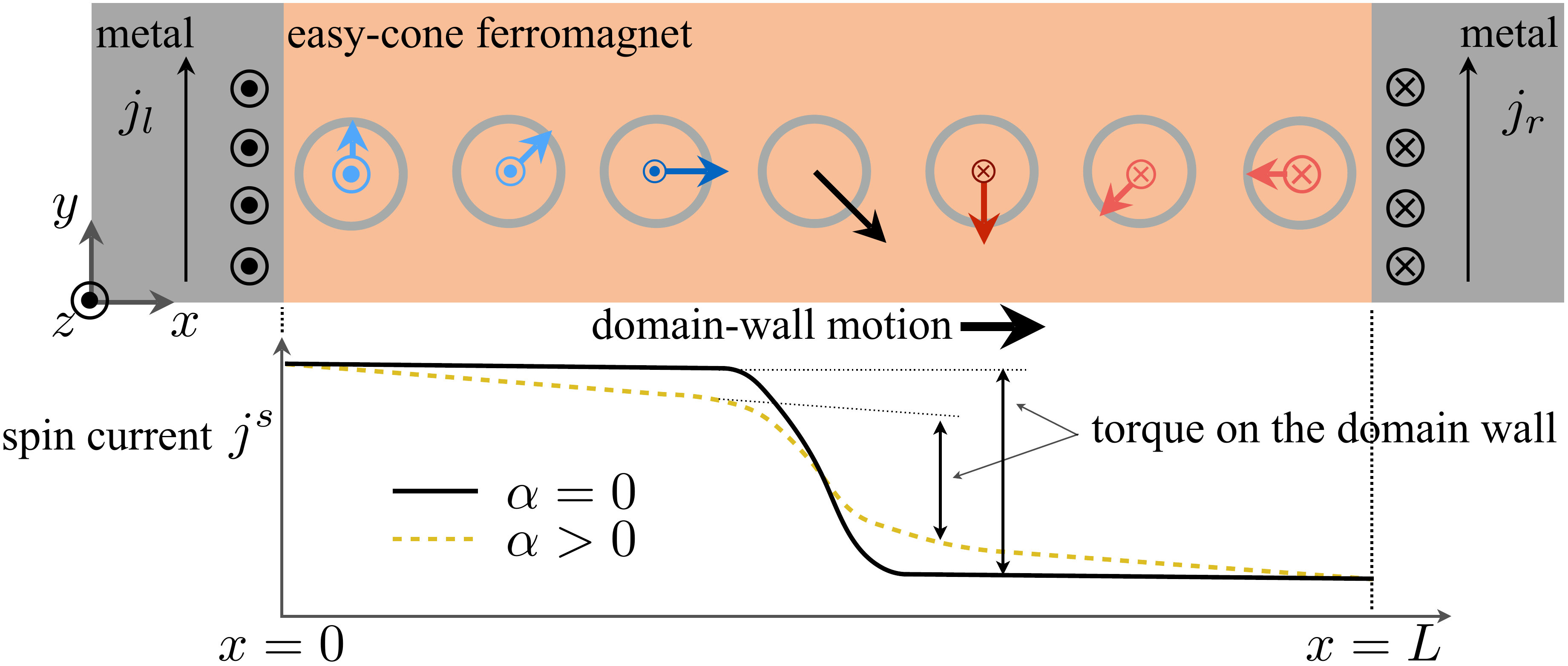}
\caption{(color online) A schematic geometrical setup for a domain-wall motion driven by spin supercurrent. Blue and red colors represent positive and negative spin component projected onto $z$ axis, respectively. The black arrow is the spin direction at the center of the domain wall. The charge currents through two adjacent heavy metals inject spin current into the magnet via spin Hall effect. The polarization directions of injected spin are depicted as circles in the metals. The spin is transported to the domain wall by spin supercurrent in the form of spiraling spin texture. The domain wall moves by absorbing the transported spin.}
\label{fig:fig3}
\end{figure}

Let us present here one of our main findings, a domain-wall motion driven by spin supercurrent. See Fig.~\ref{fig:fig3} for the schematic geometrical setup. The source and drain of spin are realized at the left and right boundaries by sandwiching the magnet with heavy metals such as Pt or Ta. Via spin Hall effect \cite{SinovaRMP2015}, the 2D charge current density $j_l$ in the left metal injects the spin current (polarized along the $z$ axis) $j^s_l = \vartheta j_l \sin^2 \theta_c$ into the ferromagnet \cite{*[][{, and references therein.}] TserkovnyakRMP2005}, where the coefficient $\vartheta$ parametrizes the efficiency of conversion from the charge current to the spin current \cite{TserkovnyakPRB2014} and $\theta_c$ is the angle that the easy cones make with the symmetry axis. Likewise, the charge current density $j_r$ in the right metal injects the spin current $j^s_r = - \vartheta j_r \sin^2 \theta_c$ into the magnet. We shall show below that the domain wall absorbs the injected spin current transported by spin supercurrent, and thereby moves at the velocity
\begin{equation}
v = \frac{g}{g^2 + \alpha^2 \eta_v \eta_\omega} \sin^2 \theta_c \, \vartheta (j_l - j_r) \, . \\
\label{eq:v}
\end{equation}
Here, $g \equiv 2 s \cos \theta_c$ is the gyrotropic coupling constant between the translational motion of the domain wall and the global spin precession about the $z$ axis \cite{ThielePRL1973}, where $s$ is the scalar spin density per unit volume;  $\alpha$ is the Gilbert damping constant; $\eta_v$ and $\eta_\omega$ are the coefficients that characterize energy dissipation associated with the linear dynamics of the domain wall and the global precessional dynamics of spins, respectively, whose explicit definitions will be given later. In the absence of damping, all of the spin current is transported by spin superfluid to the domain wall, which in turn moves at the velocity $v = \sin^2 \theta_c \, (j_l - j_r) / 2 s \cos \theta_c$ as a consequence of the conservation of spin angular momentum. Finite damping causes partial loss of spin due to the spin precession ($\propto \eta_\omega$) and the domain-wall motion ($\propto \eta_v$), which decreases the domain-wall speed.

\textcite{UpadhyayaarXiv2016} including us recently showed that spin supercurrent flowing through an easy-plane magnet can drive a domain wall in an easy-axis magnet, when two magnets are exchange-coupled. In the proposal, the domain-wall speed increases as the spin current increases in the linear regime. There is, however, a critical spin current that is proportional to the exchange-coupling strength, above which the domain-wall speed decreases significantly by entering the nonlinear regime. Differing from that, in easy-cone magnets, the domain-wall speed keeps increasing linearly as spin current increases without any breakdown as long as superfluid spin transport is stable \footnote{The velocity of spin in superfluid spin transport, which is the ratio of spin supercurrent to spin density, cannot be larger than the speed of spin waves according to the Landau criterion \cite{SoninAP2010}.}.

\emph{Easy-cone magnets.}|Our model system is a quasi one-dimensional ferromagnet with easy-cone anisotropy. When the ambient temperature is much below than the magnetic ordering temperature, the state of the system is described by the unit vector $\hat{\mathbf{n}}$ along the local spin density $\mathbf{s} \equiv s \hat{\mathbf{n}}$. It is convenient to parametrize $\hat{\mathbf{n}}$ in spherical coordinates $\theta$ and $\phi$ for our discussions: $\hat{\mathbf{n}} \equiv (\sin \theta \cos \phi, \sin \theta \sin \phi, \cos \theta)$. The potential energy of the system is given by
\begin{equation}
U = \int dV \left[ A \{ (\boldsymbol{\nabla} \theta)^2 + \sin^2 \theta (\boldsymbol{\nabla} \phi)^2 \} / 2 + K f(\theta) \right] \, ,
\label{eq:U}
\end{equation}
where $A$ and $K$ parametrize the spin-direction stiffness and the easy-cone anisotropy, respectively, and the high symmetry axis is defined as the $z$ axis. Here, a dimensionless function $f(\theta)$ is arbitrary except the following conditions: it is invariant under $n_z \mapsto - n_z$, i.e., $f(\pi - \theta) = f(\theta)$, and it attains the local minimum only at two points $0 < \theta_c < \pi / 2$ and $\pi - \theta_c$. Without loss of generality, it is assumed that the anisotropy energy vanishes at the minimum points, e.g., $f(\theta_c) = 0$. A class of the functions given by $f(\theta) = (\sin^2 \theta - \sin^2 \theta_c)^2$ will be used when providing a concrete example. The characteristic length and energy-density scales of the problem are $\sqrt{A/K}$ and $\sqrt{A K}$, respectively, in which we shall work henceforth. When the ferromagnet is narrow compared to the characteristic length scale, variations of the order parameter across the ferromagnet can be neglected: $\theta(\mathbf{r}, t) = \theta(x, t)$ and $\phi(\mathbf{r}, t) = \phi(x, t)$.

The system has two symmetries: the spin-reflection symmetry through the $xy$ plane, $\theta (x, t) \mapsto \pi - \theta (x, t)$, and the spin-rotational symmetry about the $z$ axis, $\phi (x, t) \mapsto \phi (x, t) + \delta \phi$, which we shall refer to as Z$_2$ and U(1) symmetries, respectively. The ground-state manifolds are two cones in the unit sphere that make the angle $\theta_c$ with the $z$ axis [see Fig.~\ref{fig:fig1}(c)].  A ground state lies in one of the two cones, whereby breaks the Z$_2$ symmetry; it takes an arbitrary azimuthal angle $\phi$, whereby breaks the U(1) symmetry.

Our system has two disconnected ground-state manifolds, and thus can harbor a domain wall interpolating the two ground states \cite{Kardar}. It is an extremum of the potential energy, which satisfies
\begin{subequations}
\label{eq:dU}
\begin{align}
& \delta U / \delta \theta = - \theta'' + \sin \theta \cos \theta \phi'^2 + \partial_\theta f = 0 \, , \\
& \delta U / \delta \phi = - (\sin^2 \theta \phi') ' = 0 \, ,
\end{align}
\end{subequations}
with the boundary condition $\theta (x = -\infty) = \theta_c$ and $\theta (x = \infty) = \pi - \theta_c$. The equilibrium domain-wall solution is implicitly given by
\begin{equation}
x - x_0 = \int_{\pi/2}^{\theta_0(x)} \frac{d \theta}{\sqrt{2 f(\theta)}} \, , \quad \phi (x) \equiv \phi_0 \, ,
\end{equation}
where $x_0$ is the center of the domain wall and $\phi_0$ is an arbitrary reference angle. The solution $\theta_0(x)$ can be explicitly obtained for certain cases. For example, when $f(\theta) = (\sin^2 \theta - 1/2)^2$, we have $\theta_0(x) = \pi - \arctan [\coth(x/2)]$. The explicit solution for $\theta_0 (x)$ is not necessary for our main discussion on the interaction between a domain wall and spin supercurrent, which shall be shown later, and thus we content ourselves with the implicit solution here.

Next, our system has the ground-state manifold with U(1) spin-rotational symmetry, and thus can support spin supercurrent. To discuss the dynamic steady-state that carries finite spin supercurrent, let us employ the Landau-Lifhistz-Gilbert (LLG) equations:
\begin{subequations}
\label{eq:llg}
\begin{align}
& -s \sin \theta \, \dot{\phi} - \alpha s \dot{\theta} = - \theta'' + \sin \theta \cos \theta \, \phi'^2 + \partial_\theta f \, , \label{eq:llg-theta} \\
& s \sin \theta \, \dot{\theta} - \alpha s \sin^2 \theta \, \dot{\phi} = - (\sin^2 \theta \, \phi') ' \, .
\end{align}
\end{subequations}
The latter equation in the absence of damping $\alpha = 0$ can be interpreted as the continuity equation of the spin angular momentum projected onto the $z$ axis: the time evolution of the spin density, $s \dot{n}_z = - s \sin \theta \, \dot{\theta}$, and the divergence of the spin current density, $j^s \equiv - \sin^2 \theta \, \phi'$, add up to zero. We shall set $s = 1$ hereafter by using $s/K$ as the unit time. We are interested in the nonequilibrium steady state close to the ground state with the constant polar and azimuthal angles $\theta(x) \equiv \theta_c$ and $\phi' \equiv 0$, and thus we expand the LLG equations to the linear order in $\delta_\theta \equiv \theta(x, t) - \theta_c, \phi'$, and $\dot{\phi}$, which results in
\begin{subequations}
\label{eq:l-llg}
\begin{align}
& - \sin \theta_c \, \dot{\phi} - \alpha \dot{\delta}_\theta = \kappa \delta_\theta \, , \\
& \sin \theta_c \, \dot{\delta}_\theta - \alpha \sin^2 \theta_c \, \dot{\phi} = - \sin^2 \theta_c \, \phi'' \, ,
\end{align}
\end{subequations}
where $\kappa \equiv \partial_\theta^2 f (\theta_c)$ parametrizes the curvature of the anisotropy at the local minimum point $\theta = \theta_c$. The spin current can be injected by sandwiching the magnet with heavy metals (see Fig.~\ref{fig:fig2} for the schematic geometrical setup), the effects of which can be captured by the following boundary conditions for the spin current density (projected onto the $z$ axis):
\begin{subequations}
\label{eq:bc}
\begin{align}
j^s (0) & = \sin^2 \theta(0) [\vartheta j_l - \gamma \dot{\phi}(0)] \, , \\
j^s (L) &= \sin^2 \theta(L) [\vartheta j_r + \gamma \dot{\phi}(L)] \, ,
\end{align}
\end{subequations}
within the linear response \cite{TakeiPRL2015}. Here, $\vartheta$ is the coefficient parametrizing the dampinglike torque on the magnet due to the charge current at the interfaces, which is related to the effective interfacial spin Hall angle $\Theta$ via $\vartheta = \hbar \tan \Theta / 2 e t$ with $t$ the thickness of the metals and $-e$ the charge of electrons; $\gamma \equiv \hbar g^{\uparrow \downarrow} / 4 \pi$ parametrizes the spin pumping at the interfaces with $g^{\uparrow \downarrow}$ the effective interfacial spin-mixing conductance \cite{TserkovnyakPRB2014}. The steady-state solution to the linearized LLG equations~(\ref{eq:l-llg}) with the above boundary conditions is given by
\begin{subequations}
\begin{align}
& \dot{\phi} (x, t) \equiv \omega = \frac{\vartheta (j_l - j_r)}{2 \gamma + \alpha L} \, , \\
& j^s (x, t) = \sin^2 \theta_c \left[ \vartheta j_l - (\gamma + \alpha x) \omega \right] \, ,
\end{align}
\end{subequations}
with the uniform polar angle $\delta_\theta (x, t) \equiv - \sin \theta_c \, \omega / \kappa$. Note that, by taking the limit $\theta_c \rightarrow \pi / 2$, we can recover the result for the global spin-precession frequency in the case of easy-plane ferromagnets \cite{TakeiPRL2014}.

\emph{Domain-wall motion.}|With the understanding of the physical manifestation of the broken Z$_2$ symmetry---a domain wall---and that of the broken U(1) symmetry---spin superfluidity---now let us turn to our main interest: the interaction between a domain wall and spin supercurrent. First, we study the motion of a domain wall driven by spin supercurrent. See Fig.~\ref{fig:fig3} for illustration. Specifically, we look for a steady-state solution to the LLG equations~(\ref{eq:llg}) that contains a domain wall moving at the velocity $v$ within the linear response regime. To that end, we go to the frame moving at the velocity $v$, which can be implemented by replacing $\partial_t$ by $\partial_t - v \partial_x$ in the lab-frame LLG equations~(\ref{eq:llg}). To the linear order in $v, \dot{\phi}(x, t) \equiv \omega$, and $\phi'$, the resultant LLG equations are
\begin{subequations}
\begin{align}
& - \sin \theta \, \omega + \alpha \sin \theta \, \theta' v = - \theta'' + \partial_\theta f \, , \\
& - \sin \theta \, \theta' v - \alpha \sin^2 \theta \, \omega = - (\sin^2 \theta \, \phi')' \, . \label{eq:l-llg2-phi}
\end{align}
\end{subequations}

To obtain the equations for $v$ and $\omega$, we multiply the former equation by $\theta'$ and integrate both equations over the spatial dimension, which results in
\begin{subequations}
\label{eq:vw}
\begin{align}
& - g \omega + \alpha \eta_v v = 0 \, , \label{eq:f} \\
& g v + \alpha \eta_\omega \omega = \left[ \sin^2 \theta_c \, \phi' \right]^{x = L}_{x = 0} \, , \label{eq:t}
\end{align}
\end{subequations}
to the linear order in $\delta_\theta (0) \equiv \theta(0) - \theta_c$ and $\delta_\theta (L) \equiv \theta(L) - (\pi -\theta_c)$ (which turned out to not appear in the result). Here, $g \equiv 2 \cos \theta_c$ is the gyrotropic coupling constant between $v$ and $\omega$ \cite{ThielePRL1973}, and 
\begin{subequations}
\begin{align}
& \eta_v \equiv \int_{\theta_c}^{\pi - \theta_c} d \theta \sqrt{2 f(\theta)} \, , \\
& \eta_\omega \equiv \sin^2 \theta_c \, L + \int_{\theta_c}^{\pi - \theta_c} d\theta \, \frac{\sin^2 \theta - \sin^2 \theta_c}{\sqrt{2 f(\theta)}} \, ,
\end{align}
\end{subequations}
parametrize energy dissipation associated with $v$ and $\omega$, respectively. In deriving these results, we used $\theta_0' (x) = \sqrt{2 f [\theta_0(x)]}$ to change the integration over the spatial variable $x$ to the one over the angle variable $\theta$. The latter equation~(\ref{eq:t}) represents the conservation of the spin angular momentum. The right-hand side is the net injection of the spin angular momentum into the magnet, $j^s (0) - j^s (L)$. The addition of the spin angular momentum translates into the motion of the domain wall, $g v$. The Gilbert damping causes partial loss of the spin, $\alpha \eta_\omega \omega$, which is proportional to the global precession frequency. The former equation represents the absence of a force on the domain wall. By solving Eqs.~(\ref{eq:vw}) subjected to the boundary conditions~(\ref{eq:bc}), which is invariant under the transformation $x \mapsto x - vt$ within the linear response, we obtain the self-consistent solution for $v$ and $\omega$:
\begin{subequations}
\begin{align}
v &= \frac{g}{g^2 + \alpha^2 \eta_v \eta_\omega} \sin^2 \theta_c \, \vartheta (j_l - j_r) \, , \\
\omega &= \frac{\alpha \eta_v}{g^2 + \alpha^2 \eta_v \eta_\omega} \sin^2 \theta_c \, \vartheta (j_l - j_r) \, .
\end{align}
\end{subequations}
This is our first main result. See Fig.~\ref{fig:fig3} for the schematic plot for the spatial profile of the spin current $j^s$, whose rapid drop in the domain wall represents the spin-transfer torque from the spin current $j^s$ to the domain wall.

\emph{Spin-current generation.}|Next, as a reciprocal phenomenon, we study spin-current generation by the field-induced domain-wall motion. An external magnetic field in the $z$ direction engenders a Zeeman term in the potential energy, $- h \int dx \, \cos \theta$, which creates an extra term, $h \sin \theta$, in the right-hand side of the LLG equation~(\ref{eq:llg-theta}). The modified equations for $v$ and $\omega$ are given by
\begin{subequations}
\label{eq:vw-2}
\begin{align}
& - g \omega + \alpha \eta_v v = 2 h \cos \theta_c \, ,  \\
& g v + \alpha \eta_\omega \omega = 0 \, ,
\end{align}
\end{subequations}
in the absence of the charge current in the attached metals. Here, $2 h \cos \theta_c$ is the force on the domain wall. Solving the above equations with the boundary conditions (\ref{eq:bc}), we obtain 
\begin{subequations}
\begin{align}
v &= \frac{\alpha \eta_\omega}{g^2 + \alpha^2 \eta_v \eta_\omega} 2 h \cos \theta_c \, , \\
\omega &= - \frac{g}{g^2 + \alpha^2 \eta_v \eta_\omega} 2 h \cos \theta_c \, .
\end{align}
\end{subequations}
The dynamics of spins at the interface injects spin current into the adjacent metals via spin pumping \cite{TserkovnyakRMP2005}. The amount of spin injected into the left and right metals are equal and given by
\begin{equation}
- j^s (0) = j^s (L) = \gamma \sin^2 \theta_c \, \omega \, ,
\end{equation}
which can be inferred by measuring induced charge current in the metals via inverse spin Hall effect \cite{SinovaRMP2015}. This is our second main result.

\begin{figure}
\includegraphics[width=0.8 \columnwidth]{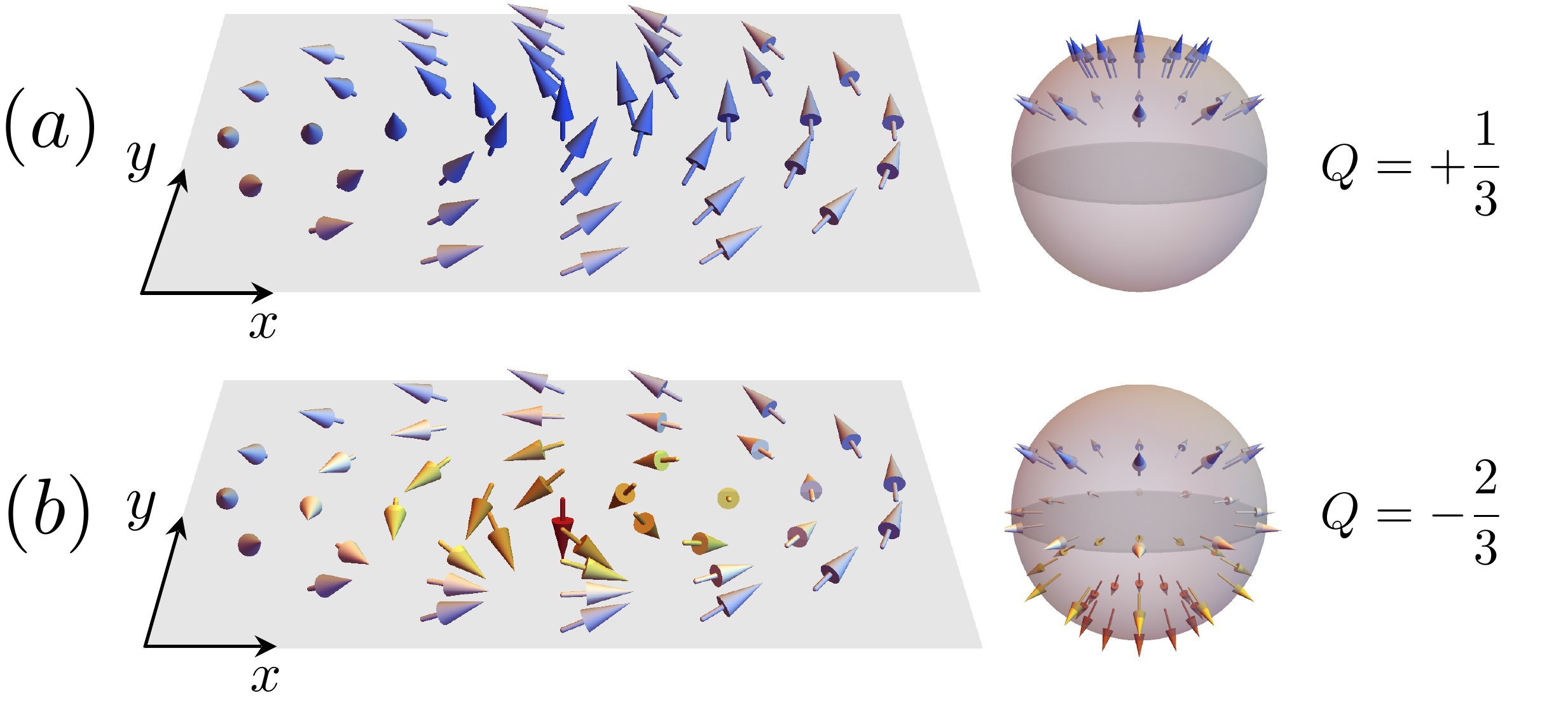}
\caption{(color online) Schematic illustrations of two types of vortices. Spins are in the upper cone away from the vortex centers. $Q$ is the Skyrmion charge of a vortex. See the main text for discussions.}
\label{fig:fig4}
\end{figure}

\emph{Discussion.}|Let us discuss how easy-cone anisotropy can arise with an example of bilayer Co/Pt \cite{DienyEPL1994}. The anisotropy energy of the system can be effectively written as $\kappa_1 \sin^2 \theta + \kappa_2 \sin^4 \theta$. The coefficient of the first term is positive, $\kappa_1 > 0$, when the cobalt film is so thin (e.g., 0.5nm thick) that the term is dominated by the interfacial easy-axis anisotropy. It becomes negative, $\kappa_1 < 0$, due to the easy-plane shape anisotropy, when the cobalt film is thick enough to have negligible interface effect. The second term comes from the bulk crystalline anisotropy and its coefficient remains positive, $\kappa_2 > 0$, independently of the cobalt thickness. When the thickness is tuned to satisfy $-2 \kappa_2 < \kappa_1 < 0$, the cobalt has easy-cone anisotropy with the canting angle $\theta_c = \arcsin \sqrt{|\kappa_1| / 2 \kappa_2}$. For example, when the Co and Pt thicknesses are 0.7nm and 1.5nm, respectively, the coefficients are $\kappa_1 = - 30$ kJ/m$^3$ and $\kappa_2 = 120$ kJ/m$^3$, which yields the equilibrium cone angle $\theta_c = 20^\circ$ \cite{SticklerPRB2011}.

To make a simple quantitative estimate for the domain-wall speed induced by spin supercurrent, let us take the following parameters: the saturation magnetization density $M_s \sim 10^6$A/m and the equilibrium cone angle $\theta_c \sim 20^\circ$ measured in bilayer Co$_\text{0.7nm}$/Pt$_\text{1.5nm}$ \cite{SticklerPRB2011, BaratiPRB2014}, and the spin Hall angle $\Theta \sim 0.1$ measured in YIG/Pt interfaces \cite{HahnPRB2013}. Then, the 2D charge current densities $j_l= 10^5$A/m and $j_r = 0$ through the $5$nm-thick platinums will yield the domain-wall speed of $v \sim 7$m/s, when neglecting the Gilbert damping.

We have studied the interaction between a domain wall and spin supercurrent in quasi-one-dimensional easy-cone ferromagnets. Coexistence of spin superfluidity and a domain wall can lead to other possibly interesting phenomena. For example, two-dimensional magnets with easy-cone anisotropy support vortices, topological defects associated with U(1) symmetry, which can interact with a domain wall. Since vortices cause phase slips disturbing spin supercurrent \cite{SoninAP2010}, their interaction may have an interesting effect on phase-slip-induced resistances of spin supercurrent \cite{KimPRB2016, *KimPRL2016}. In addition, easy-cone magnets support two types of magnetic vortices, which have different Skyrmion-charge magnitudes due to the broken Z$_2$ symmetry. See Fig.~\ref{fig:fig4} for illustrations. These Skyrmion charges have important effects on the dynamics of vortices by determining the gyrotropic coupling between two spatial coordinates \cite{ThielePRL1973}. Vortices in easy-cone magnets will thus show the two distinct gyrotropic dynamics, which cannot be observed in easy-plane magnets that can only support vortices with the Skyrmion charges of the same magnitude, $Q = \pm 1/2$. 

Some geometrically frustrated magnets such as the Heisenberg antiferromagnets on the Kagome lattice also have the ground states characterized by Z$_2$ $\times$ U(1) \cite{*[][{, and references therein.}] StarykhRPP2015, *[][{, and references therein.}] KorshunovPU2006}, which are associated with two possible chiralities of spin configuration and spin rotations about the global symmetry axis. We envision that, if the chirality can be coupled to the net spin density by, e.g., engineering a certain spin-orbit coupling, it would be possible to drive a domain wall connecting two chiralities by spin supercurrent via the mechanism discussed in the Letter.

\begin{acknowledgments}
We thank Daniel Hill and Pramey Upadhyaya for useful discussions. This work was supported by the Army Research Office under Contract No. 911NF-14-1-0016 and, in part, by FAME (an SRC STARnet center sponsored by MARCO and DARPA).
\end{acknowledgments}

\bibliographystyle{/Users/evol/Dropbox/School/Research/apsrev4-1-nourl}
\bibliography{/Users/evol/Dropbox/School/Research/master}

\end{document}